\documentclass[preprint,eqseqnum,12pt]{elsarticle}

\usepackage{epsfig}
\usepackage{amssymb}
\usepackage{amsmath}
\usepackage{lineno}
\biboptions{sort&compress}

\usepackage{bm}
\usepackage{amsmath}
\usepackage[utf8]{inputenc}

\numberwithin{equation}{section}

\begin{document}

\begin{frontmatter}




\title{The Mass Gap Approach to QCD. \\ III. Solution of the gluon confinement problem}



\author{Vakhtang Gogokhia}
\ead{gogohia.vahtang@wigner.hu}
\author{Gergely G\'abor Barnaf\"oldi}
\ead{barnafoldi.gergely@wigner.hu}

\address{Wigner Research Centre for Physics\\
29-33 Konkoly-Thege Mikl\'os Str., H- 121, Budapest, Hungary}

\begin{abstract}
We present reports on progress in the solution of the gluon confinement problem and in the explanation of the scale breaking in the asymptotic freedom regime in QCD. 
It is based on the previously formulated the mass gap approach to QCD, which sheds the light on a mass origin in its ground state. 
We have established the exact expression for the full gluon propagator. One of its solution allows to sum up all the severe (more singular than $1/q^2$) infrared singularities which can be accommodated
by a single full gluon propagator. In order to correctly treat them, we have developed a formalism of the joint use of the theory of functions of complex variable, and the dimensional  
regularization method correctly implemented into the theory of distributions. Despite the presence of the mass gap, gluons remain massless. Our solution satisfies the two boundary conditions: it dominates over all other possible solutions at large distances but itself prevents gluons to appear in the physical spectrum. It possesses the asymptotic freedom behaviour at short distances.  
We discuss the connection of our results to the Elitzur's theorem.
\end{abstract}



\end{frontmatter}

\hspace{2mm}

\section{Introduction. The mass gap approach to QCD}

\hspace{2mm}

In order to calculate the physical observables in Quantum Chromodynamics (QCD) from first principles, we need the full gluon propagator, which should satisfy the two boundary conditions:
confinement when gluon momentum goes to zero $q^2 \rightarrow 0$ (large distances) and asymptotic freedom (AF)~\cite{1,2} when it goes to infinity $q^2 \rightarrow \infty$ (short distances). 
In our previous work~\cite{3} (which is the first paper of the three paper series announced there)
a new expression for the full gluon propagator has been found. For this the formalism of the mass gap approach to QCD has been developed. It is worth underlining that it has been formulated 
without making any kind of the simplifications and using the tensor algebra derivation rigorous rules only.

Let us now show up the new expression for the regularized full gluon propagator obtained in~\cite{3} within the mass gap approach to QCD, namely  

\begin{equation}
D_{\mu\nu}(q) =  { i T_{\mu\nu}(q) \over q^2 + q^2 \Pi(q^2) + \Delta^2_t(D) }  + i L_{\mu\nu}(q) { \lambda^{-1} \over q^2 + \lambda^{-1} \Delta^2_t(D)},
\end{equation}
where we have introduced the useful notation $\lambda^{-1}$~\cite{3,4} for the gauge-fixing parameter of the free gluon state. It can be obtained from eq.~(1.1) in the 
formal $\Pi(q^2) = \Delta^2_t(D) = 0$ limit and denoted as $D^0_{\mu\nu}(q)$, see eq.~(2.18) below.

The invariant function of the full gluon propagator is defined as follows:

\begin{equation}
d(q^2) = {1 \over 1 + \Pi(q^2) + (\Delta^2_t(D) / q^2)}.
\end{equation}

In both expressions the invariant function $\Pi(q^2) =\Pi^s(q^2; D)$  is regular at zero and only logarithmic divergent in the perturbation theory (PT) $q^2 \rightarrow  \infty$
limit. The only one quadratically ultra-violet (UV) divergent constant, but regularized from above and below,
present explicitly in eq.~(1.1), is the so-called tadpole term $\Delta^2_t(D)$, having the dimension of a mass squared. In~\cite{3}  it has been exactly proven that it cannot be removed from the 
QCD ground state by any means. However, its formal $\Delta^2_t(D) = 0$ limit exists, and QCD in this limit is known as PT QCD~\cite{3}, and its full gluon propagator, denoted as $D^{PT}_{\mu\nu}(q)$, 
can be obtained from eq.~(1.1) in the above-mentioned formal limit, see eq.~(2.17) below, while $\Pi^s(q^2; D)$ becomes $\Pi^s(q^2; D^{PT})$.

The renormalized version of the tadpole term $\Delta^2 = Z^{-1}_{\Delta} \Delta^2_t(D)$, introduced in~\cite{3}, will be conventionally called a mass gap, and   
it is finite and positive, by definition. The non-perturbative (NP) renormalization constant $Z_{\Delta} = Z_{\Delta}(\lambda) = Z(\lambda)$ is a function of the dimensionless 
regulating parameter  $\lambda$ (do not mixed it up with the gauge-fixing parameter $\lambda^{-1}$,  introduced above). For the description of the transcendental character of the tadpole term and its renormalization see the corresponding appendix D.

The ST identity, as it follows from eq.~(1.1),  now becomes

\begin{equation}
q_{\mu}q_{\nu} D_{\mu\nu}(q) = i \xi(q^2)  = i { \lambda^{-1} q^2 \over q^2 + \lambda^{-1} \Delta^2_t(D)}.
\end{equation}
Here $\xi(q^2)$ is the gauge-fixing parameter of the full gluon propagator and its solution in the right-hand side is called  as the generalized gauge since it depends on 
the tadpole term $\Delta^2_t(D)$. When it is zero, one recovers the gauge-fixing parameter for the free gluon propagator.
To our best knowledge the generalized gauge (1.3) has been derived in~\cite{3} (and see references therein). 

Concluding, the whole gluon momentum range is $q^2 \in [0, \infty)$. In this paper we work in Euclidean metric $q^2=q^2_0 + \bf q^2$ since
it implies $q_i \rightarrow 0$ when $q^2 \rightarrow 0$ and {\it vice versa}. This makes it possible to avoid the unphysical infrared (IR) singularities on the light cone.
 As usual in Euclidean metric in eq.~(1.1) $T_{\mu\nu}(q) = \delta_{\mu\nu} - q_{\mu}q_{\nu} / q^2 = \delta_{\mu\nu} - L_{\mu\nu}(q)$.

\hspace{2mm}

\section{Singular solution}

\hspace{2mm}

Let us begin with the remembering that the expression (1.2) is not a solution for the full gluon invariant function,
but being rather some kind of the NL transcendental equation for the different invariant functions $d(q^2), \ \Pi(q^2; d)$ and the tadpole term $\Delta^2_t(d)$,
i.e., $d=f(d)$. It can be equivalently written down as follows:

\begin{equation}
d(q^2) = 1 - \left[ \Pi(q^2; d) + { \Delta^2_t(d) \over q^2} \right] d(q^2),
\end{equation}
indeed. Note, here and below we equivalently replaced $D$ by $d$, and thus in the formal $\Delta^2_t(d) = 0$ limit $d$ has to be replaced by $d^{PT}$.
Its (non-linear) NL iteration solution has been developed in~\cite{5}.
However, here we are going to present much more economic and mathematically transparent method of its solution. It leads to the same result as it present in the 
iteration solution~\cite{5}, as it needs to be.

For this purpose, let us re-write the initial expression (1.2) in the slightly different but equivalent way, namely

\begin{equation}
d(q^2; z) = {1 \over 1 + \Pi(q^2; d) + z Z_{\Delta}}, \quad z = { \Delta^2 \over q^2},
\end{equation}
i.e., we have already used the above-mentioned definition  $\Delta^2_t(D) = Z_{\Delta}\Delta^2$,  in order to introduce the dependence on the dimensionless variable $z$, i.e., on 
the mass gap $\Delta^2$, which is finite and positive, by definition. Since the invariant 
function $\Pi(q^2; d)$ is regular function at small $q^2$, it can not depend on $z$, while may still depending on the tadpole term or, equivalently, the mass gap itself. Let us underline that within the mass gap approach whatever is the dependence of $\Pi(q^2; d)$ on the mass gap (even can be non-analytic due the complicate transcendental character of the relation (2.2)) the formal $\Delta^2=0$ limit for this invariant function always exists and it is $\Pi(q^2; d^{PT})$, i.e., in this limit $d$ is to be replaced with $d^{PT}$, as mentioned above.
Then expression (2.2) can be formally expanded in the Taylor series in powers of $z$ around zero $z$, i.e., in fact around small $\Delta^2$, namely

\begin{equation}
d(q^2; z) = \sum_{k=0}^{\infty} z^k f_k(q^2),
\end{equation}
where the functions $f_k(q^2)$ are the corresponding derivatives of $d(q^2; z)$ with respect to $z$ and putting after $z=0$, which implies the
formal $\Delta^2=0$ limit. This means that $d(q^2; z=0)= d^{PT}(q^2)$ as well.
Thus, these invariant functions are

\begin{equation}
f_k(q^2) = (-1)^k d^{PT}(q^2)[d^{PT}(q^2) Z_{\Delta}]^k
\end{equation}
and

\begin{equation}
d(q^2; z=0) = d^{PT}(q^2) = f_0(q^2) = { 1  \over  1 + \Pi(q^2; d^{PT})},
\end{equation}
as it follows from eq.~(2.2).
It is important to emphasize that contrary to expression (2.2), expansion (2.3) can be considered now as a formal solution for
$d(q^2)$, since the invariant functions $f_k(q^2)$ depend on $d^{PT}(q^2)$, which is assumed to be known.
Expansion (2.3) then can be present as follows:

\begin{equation}
d(q^2; z) = \sum_{k=0}^{\infty} z^k f_k(q^2) = d^{PT}(q^2) + d^{INP}(q^2; z),
\end{equation}
where

\begin{equation}
d^{INP}(q^2; z) = \sum_{k=1}^{\infty} z^k f_k(q^2) = z \sum_{k=0}^{\infty} z^k f_{k+1}(q^2),
\end{equation}
and the superscript (INP) means intrinsically NP (for the explanation see section 3 below).  
The invariant functions $f_{k+1}(q^2)$ are regular functions of their argument, since they are expressed in terms of $d^{PT}(q^2)$
through the relations (2.4), and $d^{PT}(q^2)$ itself is regular function of $q^2$. Each of these functions
can be present as series in inverse powers of the variable $z$, namely $z^{-1} = (q^2 / \Delta^2)$, i.e., in powers of small $q^2$,
and the finite mass gap $\Delta^2$, which can be understood as the scale of the INP dynamics in the deep IR region (see discussions below).  
Then one obtains

\begin{equation}
f_{k+1}(q^2) = \sum_{n=0}^{\infty}  z^{-n} f^{(n)}_{k+1}(0),
\end{equation}
and substituting this series into the expansion (2.7), one arrives at

\begin{equation}
d^{INP}(q^2; z) = z \sum_{k=0}^{\infty} z^k f_{k+1}(q^2) = z \sum_{k=0}^{\infty} z^k \sum_{n=0}^{\infty}  z^{-n} f^{(n)}_{k+1}(0).
\end{equation}

After doing some algebra, this double sum series can be present as the sum of the two separate series as follows:

\begin{equation}
d^{INP}(q^2; z) = z \sum_{k=0}^{\infty} z^k \Phi_k(0) + z \sum_{k=1}^{\infty} z^{-k} \Phi_{-k}(0),
\end{equation}
where the coefficients $\Phi_k(0)$ and $\Phi_{-k}(0)$ by themselves are infinite series expressed in terms of the coefficients in expansion (2.9).
The sum of these two series in eq.~(2.10) is nothing else but the Laurent expansion, and going now back to the variable $q^2$, one gets

\begin{equation}
d^{INP}(q^2) = \Bigl( {\Delta^2 \over q^2} \Bigr) L (q^2), 
\end{equation}
with

\begin{equation}
L(q^2) = \sum_{k= - \infty}^{\infty} \Bigl(
{\Delta^2 \over q^2} \Bigr)^k \Phi_k(0), \quad   r < {\Delta^2 \over q^2} < R,
\end{equation}
and the arbitrary finite numbers $R, r$ determine the ring of convergence of the corresponding Laurent expansion (let us remind that within the singular solution the variable $ z = \Delta^2 / q^2$ is always finite). The coefficient 'functions' (or simply the coefficients in some places above and below)
\begin{equation}
\Phi_k (0) \equiv \Phi_k(\lambda, \alpha,  g^2, ...) = \sum_{m=0}^{\infty} \Phi_{km}(\lambda, \alpha, g^2, ...)
\end{equation}
may, in general, depend on all the possible non-physical parameters of the theory, for example such as the dimensionless UV regulating parameter $\lambda$, arbitrary subtraction point 
$\alpha$, $Z_{\Delta}$, etc. Let us remind that these constant functions are fragments of the skeleton loop integrals, contributing to the full gluon self-energy after subtracting from them the UV divergent
terms and then differentiate them with respect to $z$, as described above.  
The dependence of these coefficient functions on the coupling constant $g^2$, in general, is not known, so that series (2.12), being the Laurent expansions, at the same time are the
so-called ''cluster'' expansions, introduced in~\cite{6}. Also, let us note that the sum over
index $m$ in eq.~(2.13) shows that the infinite number of the contributions invoke each term in powers of the variable $(\Delta^2 / q^2)$
in the Laurent expansion (2.12). Thus, the functional dependence of the INP invariant function $d^{INP}(q^2)$ in eq.~(2.11) is fixed up to the arbitrary
but regularized constants $\Phi_k (0)$, which appear in the corresponding Laurent expansion (2.12).

Substituting decomposition (2.6) into the full gluon propagator (1.1), and doing some algebra, one obtains

\begin{equation}
D_{\mu\nu}(q) = D^{INP}_{\mu\nu}(q) + \tilde{D}^{PT}_{\mu\nu}(q),
\end{equation}
where

\begin{equation}
D^{INP}_{\mu\nu}(q) = i T_{\mu\nu}(q)d^{INP}(q^2) {  1 \over q^2},
\end{equation}
while $d^{INP}(q^2)$ is defined by expressions (2.11)-(2.12), and

\begin{equation}
\tilde{D}^{PT}_{\mu\nu}(q) = i T_{\mu\nu}(q)d^{PT}(q^2) {  1 \over q^2} + i L_{\mu\nu}(q) { \lambda^{-1} \over q^2 + \lambda^{-1} \Delta^2_t(D)}
\end{equation}
is the so-called spurious PT full gluon propagator,
which is legitimate to introduce, since any gluon propagator is defined up to its longitudinal component. In the PT $q^2 \rightarrow \infty$ limit, which is equivalent 
to formally put $\Delta^2_t(D) = 0$, it becomes

\begin{equation}
D^{PT}_{\mu\nu}(q) = i T_{\mu\nu}(q)  d^{PT}(q^2) {  1 \over q^2} + i \lambda^{-1} L_{\mu\nu}(q) { 1 \over q^2 },
\end{equation}
i.e., the expression for the full gluon propagator in PT QCD, and reminding~\cite{3} that its invariant function is regular at zero and may have only logarithmic divergences a t infinity. 
Let us stress that the sum in expression (2.14) is not a sum
of the corresponding asymptotics. The both terms in this sum are defined in the whole gluon momentum range as the full gluon propagator itself.
Putting formally in the previous equation  $d^{PT}(q^2) = 1$, one obtains the expression for the free gluon state, namely

\begin{equation}
D^0_{\mu\nu}(q) = i T_{\mu\nu}(q) {  1 \over q^2} + i \lambda^{-1} L_{\mu\nu}(q) { 1 \over q^2 }.
\end{equation}

\hspace{2mm}

\section{INP gluon propagator}

\hspace{2mm}

Establishing thus that the INP gluon propagator being the main dynamical part of the full singular propagator (2.14), let us now discuss its general properties.
On account of eqs.~(2.11) and (2.15), it becomes

\begin{equation}
D^{INP}_{\mu\nu}(q) = i T_{\mu\nu}(q) { \Delta^2 \over (q^2)^2} \times L(q^2).
\end{equation}
It is defined in the whole gluon momentum range, while its Laurent expansion

\begin{equation}
L(q^2) = \sum_{k = - \infty}^{\infty} \Bigl( {\Delta^2 \over q^2} \Bigr)^k \Phi_k(0), \quad \quad r <  {\Delta^2 \over q^2} < R
\end{equation}
is uniformly convergent only within its ring of convergence, shown above.
The most important and interesting features of the INP gluon propagator (3.1)-(3.2) are as follows:

\begin{itemize}
\item[A.] Transverse by construction, i.e., not a gauge choice by hand.
\item [B.] Its functional dependence on variable $(\Delta^2/q^2)$ is fixed up to the arbitrary, but regularized constants $\Phi_k(0)$.
\item[C.] The presence of severe (or, equivalently, INP) IR singularities

$(q^2)^{-2-k}, \ k=0,1,2,3,...$ only, while PT IR singularity is $(q^2)^{-1}$.
\item[D.] They cannot be summed up to a some known function, like it has been achieved for the longitudinal component of the full gluon propagator, explicitly shown in eq.~(2.16).
\item[E.] The summation of all the possible severe IR singularities with respect to the gluon momentum. So that all the QCD full vertices can be considered as regular functions  
of all the gluon momenta involved.
\item[F.] Has an essential singularities at $ 0 \leftarrow q^2 \rightarrow \infty$, i.e.,

beyond its ring of convergence.
\end{itemize}

Because of all these features pointed out in this section, namely being transverse, possesses dependence on the mass gap along with the explicit presence of severe IR singularities,
we call it as the INP one.

\hspace{2mm}

\subsection{Asymptotics of the INP gluon propagator}

\hspace{2mm}

Let now investigate the asymptotic properties of the INP part (3.1) of the full singular gluon propagator (2.14). 
In the limits of small and large gluon momenta,  they will be up to the behaviour of the Laurent expansion in eq.~(3.2) beyond its ring of convergence explicitly shown in this expansion. It is well known that the Laurent expansion, being a meromorphic function, is uniformly convergent at any point of its ring of convergence~\cite{7}. At the same time, in the limits $q^2 \rightarrow 0$ and $q^2 \rightarrow \infty$, i.e., beyond its ring of convergence, it has an essential singularities. The behaviour of any meromorphic function near its essential singularity is governed by the theorem from the theory of functions of complex variables~\cite{8}, namely

\hspace{2mm}

{\bf Picard theorem:} If $z_0$ is an essential singularity of the function $f(z)$, then for any complex/real number $Z \neq \infty$, expecting, may be,
one value $Z=Z_0$, every neighbourhood of $z_0$ contains infinite set of points $z$, such that

\begin{equation}
f(z) = Z, \quad z \rightarrow z_0.
\end{equation}


So that the Picard theorem tells us that the function $f(z)$ in the close neighbourhood of its essential singularity $z_0$ can be replaced by the finite
number $Z$, which value depends only on how precisely $z \rightarrow z_0$. 
Then from the Picard theorem it follows that $L(q^2) = Z_s$ when $q^2 \rightarrow 0$ and $L(q^2) = Z_r$ when $q^2 \rightarrow \infty$, where
$Z_s$ and $Z_r$ are the corresponding numbers, which do not depend on the non-physical parameters of the theory, shown in eq.~(2.13). Their values depend
only on the way how exactly the corresponding limits are approaching. Applying these results to the INP gluon propagator (3.1) -(3.2), one obtains

\begin{equation}
D^{INP}_{\mu\nu}(q; \Delta^2) = i T_{\mu\nu}(q) { \Delta^2 \over (q^2)^2} \times Z_s, \quad q^2 \rightarrow 0,
\end{equation}
and

\begin{equation}
D^{INP}_{\mu\nu}(q; \Delta^2) = i T_{\mu\nu}(q) { \Delta^2 \over (q^2)^2} \times Z_r, \quad q^2 \rightarrow \infty,
\end{equation}
where subscripts "s" and "r" in $Z_s$ and $Z_r$ stand for singular and regular, respectively. In accordance with the Picard theorem, we will consider in what follows only 
the finite values of such type of numbers. Then the two different renormalized constants appear in the two different regimes:

\begin{equation}
\Delta^2 \times Z_s = \Delta^2_{INP}, \quad (q^2 \rightarrow 0),
\end{equation}
and

\begin{equation}
\Delta^2 \times Z_r = \Delta^2_{AF}, \quad (q^2 \rightarrow \infty),
\end{equation}
to which can be assigned the physical meaning as determining the scales of the INP and non-trivial PT dynamics
in the AF regime in the true QCD ground
state, respectively. None of them depend on the non-physical parameters of the theory, as underlined above.
The basic element of these constants is the mass gap $\Delta^2$ itself. It is of the NP dynamical origin and comes from the IR region, as
described in~\cite{3}. The enhancement of the NP small and suppression of the PT large gluon modes shown in eqs.~(3.4)
and (3.5), respectively, with the finite corresponding constants correctly describe the true dynamical structure of the full singular gluon propagator (2.14).

Within the singular solution the mass gap $\Delta^2$ is always finite, but now we can consider the formal $\Delta^2=0$ limit in this solution.
From expression (3.1), one concludes that in the formal $\Delta^2=0$ limit the Laurent expansion (3.2) has an essential singularity.
Then according to the Picard theorem in its close neighbourhood, the Laurent expansion $ L(q^2) = L(q^2; \Delta^2)$ is to be replaced by the finite number $\tilde Z$ as follows:

\begin{equation}
D^{INP}_{\mu\nu}(q; \Delta^2) = i T_{\mu\nu}(q) {\Delta^2 \over (q^2)^2} \times L (q^2) = i T_{\mu\nu}(q) {\Delta^2 \over (q^2)^2} \times \tilde Z,
\end{equation}
and therefore it goes to zero in the $\Delta^2 \rightarrow 0$ limit. As a result, the full gluon propagator (2.14) becomes the PT gluon propagator
(2.17), as expected. This is in complete agreement with the statement about the existence of the formal $\Delta^2=0$ limit made above at the beginning of the previous section.

\hspace{2mm}

\section{Suppression of the free and PT QCD gluon states at $q^2 \rightarrow 0$}

\hspace{2mm}

The most surprising feature of the INP part of the singular solution (2.14) is that its structure at zero
($q^2 \rightarrow 0$) and at infinity ($q^2 \rightarrow \infty$), i.e., outside the ring of the uniform convergence for the Laurent expansion (3.2),
is determined by the composition $\Delta^2 /(q^2)^2$ only!, which is explicitly present in its INP part (3.1), indeed.
We will call it as the Picard effect in what follows, because of relations (3.4)-(3.7).
Let us note that in our previous publications~\cite{9,10,11,12,13} we called (3.4) as the zero modes enhancement model, since we have used it as confining ansatz for the 
structure of the full gluon propagator at $q^2 \rightarrow 0$. It leads to the linear rising potential between heavy quarks, see the corresponding appendix. In other words, we 
have paid much more attention to the functional dependence of this composition than to its dependence on the mass scale parameter.
At that time, we have had no idea about the origin of the mass scale parameter needed for it. Its importance has been realized when we have read the Jaffe and Witten (JW) theorem 
(unfortunately, rather late) after its formulation in~\cite{6}. In our first paper~\cite{3} we have proven the existence of a such mass scale parameter in the QCD ground state. 
It will survive the corresponding INP multiplicative (MP) IR renormalization program, see sections below.
The IR $\sim q^{-4}$ behaviour of the full gluon propagator in the different approximations and gauges has been detected long ago,
see for example~\cite{9,10,11,12,13}, and for more extensive list of references see~\cite{9}. Of course, there were no any connection to the mass gap. 

Now we are able to investigate the behaviour of the full singular propagator (2.14) as a function of its argument in the $q^2 \rightarrow 0$ limit, i.e., at large distances.
On account of expressions (2.14)-(2.16), it becomes

\begin{equation}
D_{\mu\nu}(q) = D^{INP}_{\mu\nu}(q) +  i T_{\mu\nu}(q)d^{PT}(q^2) {  1 \over q^2} + i L_{\mu\nu}(q) { \lambda^{-1} \over q^2 + \lambda^{-1} \Delta^2_t(D)}
\end{equation}
It is easy to show that equivalently this expression can be present as follows:

\begin{equation}
D_{\mu\nu}(q) = D^{INP}_{\mu\nu}(q) +  D^{PT}_{\mu\nu}(q) -  i L_{\mu\nu}(q) { \lambda^{-2}  \Delta^2_t(D)  \over q^2 (q^2 + \lambda^{-1} \Delta^2_t(D))},
\end{equation}
and thus the full singular gluon propagator contains the PT QCD full gluon propagator, which makes it possible to compare both gluon field configurations at the same time.  
The INP gluon propagator is given by eq.~(3.1). Due to the Picard effect, discussed in detail in the previous section, it is always convenient to factorize the dependence on $q^{-4}$
in the full gluon propagator in order to investigate its asymptotic in $q^2 \rightarrow 0$ limit. Then from eq.~(4.1) one finally obtains 

\begin{equation}
D_{\mu\nu}(q) = {i \over (q^2)^2} \Bigl [  T_{\mu\nu}(q) \Bigl ( \Delta^2_{INP} + q^2 d^{PT}(q^2) \Bigr ) +  L_{\mu\nu}  {q^4 \over \Delta^2_t(D)} \Bigr ],  \quad  q^2 \rightarrow 0,
\end{equation}
on account of eqs.~(3.1) and (3.4), (3.6), in this limit. The remarkable feature of this asymptotic is its pure transverse structure, since its longitudinal part is strongly suppressed, and both gluon states remain massless. But the most important feature is that it clearly shows
the suppression of the free ($d^0(q^2)=1$) and the PT QCD gluon states in comparison with the INP part of the full singular gluon propagator (let us remind that  $d^{PT}(q^2)$ is always regular function
of $q^2$). In other words, neither free gluon nor the PT QCD gluon states will appear in the physical spectrum at large distances ($q^2 \rightarrow 0$).
So that the full singular gluon propagator automatically becomes free from all the types of the PT contributions (''contaminations'', see remarks in appendix B). 
 Let us note that the above-mentioned suppression of all the types of the PT contributions can be also analytically achieved by the definition of the truly NP (TNP) full gluon propagator via eq.(4.2) as follows: 

\begin{equation}
D^{TNP}_{\mu\nu}(q) = D_{\mu\nu}(q)  -  D^{PT}_{\mu\nu}(q)  =  D^{INP}_{\mu\nu}(q) -  i L_{\mu\nu}(q) { \lambda^{-2}  \Delta^2_t(D)  \over q^2 (q^2 + \lambda^{-1} \Delta^2_t(D))},
\end{equation}
and in the $q^2 \rightarrow 0$ limit its behaviour will be dominated by the INP part like as in eq.~(4.3) despite of the different unimportant longitudinal parts.  The INP dynamics of eqs.~(4.2) and (4.4) are the same, which is important.
However, the problem remains how to confine to the vacuum the singular gluon states themselves. The corresponding full singular gluon propagator accommodates all the possible severe IR
singularities. Therefore, they can be only treated within the theory of distributions (generalized functions)~\cite{14}.

Concluding, let us remind that  at the finite gluon momentum the full gluon propagator (1.1) may have the pole-type singularity with exactly defined gluon pole mass.
However, this pole cannot be treated as the mass-shell singularity, i.e., requiring mass dynamically, the massive gluon cannot appear in the physical spectrum~\cite{15}.

\hspace{2mm}

\section{Theory of distributions and  dimensional regularization}

\hspace{2mm}

In order to investigate severe IR structure of the full singular gluon propagator (2.14), it is convenient to begin with some general discussion 
of the properties of the gauge particles Green's 
functions in the IR region. In principle, all the Green's functions (propagators) in QCD are generalized functions, i.e., they are distributions~\cite{14}. This is especially true
for the severe IR singularities of the full gluon propagator due to the self-interaction of massless gluons in the QCD vacuum, discussed and derived above.
They present a rather broad and important class of functions with algebraic singularities, i.e., functions with non-summable
singularities at isolated points (at zero in what follows).
Roughly speaking, this means that all the relations involving
distributions should be considered under corresponding integrals,
taking into account the smoothness properties of the corresponding
space of test functions. Let us note in advance that the
space in which our generalized functions are continuous linear
functionals is $K$, that's the space of infinitely differentiable functions having compact support,
i.e., they are zero outside some finite region (different for each differentiable function).

The INP gluon propagator (3.1), being the part of the full gluon propagator, is defined in the whole gluon momentum range.
At the same time, the corresponding Laurent expansion (3.2) is uniformly convergent only at any point of its ring of convergence.
Thus, the INP gluon propagator (3.1) contains the infinite number of severe IR singularities $\sim (q^2)^{-2-k}, \ k=0,1,2,3,...$
when $q^2 \rightarrow 0$, i.e., beyond its ring of convergence. We will show below that they cannot be treated as the standard PT IR singularity,
which is as much singular as $\sim (q^2)^{-1}$ only. The theorem, which allows to put the severe IR singularities under firm mathematical control, is as follows~\cite{14}:

\hspace{5mm}

{\bf Gelfand-Shilov (GS) theorem:} For the positive quadratic form

\begin{equation}
P(q) = q^2_0 + q^2_1 +q^2_2 +...+ q^2_{n-1} = q^2 > 0,
\end{equation}
and $n$ is even number, let us define the generalized function $P^{\lambda}$ as follows:

\begin{equation}
(P^{\lambda}, \varphi) = \int d^dq P^{\lambda}(q) \varphi(q),
\end{equation}
where $\lambda$ is a complex number and superscript $d$ is the dimension of the loop integral. The function
$\varphi(q)$ is the above-mentioned some test function. At $Re \lambda \geq 0$ this integral
is convergent and analytic function of $\lambda$. For $Re \lambda < 0$  this integral has a simple poles at points

\begin{equation}
\lambda = - {n \over 2} - k, \quad k=0,1,2,3... \ ,
\end{equation}
so that the distribution $P^{\lambda}(q)$ itself can be present as

\begin{equation}
P^{\lambda}(q) = (q^2)^{\lambda} = { C^{(k)}_{-1} \over \lambda + (d/2) +k } + finite \ terms,
\end{equation}
where residues are

\begin{equation}
 C^{(k)}_{-1} = { \pi^{n/2} \over 2^{2k} k! \Gamma((n/2) + k) } \cdot D^k \delta^n (q),
\end{equation}
while $D$ denotes d'Alembert/Laplace operator in Euclidean metric, so it is

\begin{equation}
D = { \partial^2 \over \partial q_0^2} + { \partial^2 \over \partial q_1^2} +...+ { \partial^2 \over \partial q^2_{n-1} }.
\end{equation}

{\bf end of the theorem:}

\hspace{2mm}

In the dimensional regularization method (DRM)~\cite{16} the UV divergences and IR singularities are regulated by admitting that

\begin{equation}
d = n - 2 \delta, \ \delta \rightarrow 0^+ \ \textrm{and} \quad d = n + 2 \epsilon, \quad \epsilon \rightarrow 0^+,
\end{equation}
respectively. Then from eq.~(5.4), on account of eqs.~(5.3) and (5.7), it follows that

\begin{equation}
\lambda + {d \over 2} + k = \epsilon, \quad \epsilon \rightarrow 0^+,
\end{equation}
and corresponding expression (5.4) can be now treated as the dimensional regularization expansion (DRE) in powers of $\epsilon$, namely

\begin{equation}
(q^2)^{\lambda} = { 1 \over \epsilon } C^{(k)}_{-1}  + finite \ terms , \quad \lambda = - (n / 2) - k, \ k=0,1,2,3... , \quad \epsilon \rightarrow 0^+,
\end{equation}
where we can put $d=n$ now (i.e., after introducing this
expansion). By the "$finite \ terms$" here and everywhere a number of
necessary subtractions under corresponding integrals is understood
\cite{14} (see below as well). Let us underline its most remarkable feature. The order of
singularity does not depend on $n$ and $k$ (and thus on $\lambda$). In terms of
the IR regularization parameter $\epsilon$ it is always a simple
pole $1/ \epsilon$. However, the residue at a pole will be drastically
changed from one power singularity to another. This means
different solutions to the whole system of the dynamical equations for
different set of numbers $\lambda$ and $k$. Different solutions
mean, in their turn, different vacua. In this picture different
vacua are to be labelled by the two independent numbers: the
exponent $\lambda$ and $k$. At a given number of $d(=n)$ the
exponent $\lambda$ is always negative being integer if $d(=n)$ is
an even number or fractional if $d(=n)$ is an odd number. The
number $k$ is always integer and positive and precisely it
determines the corresponding residue at a simple pole, see
eq.~(5.5). It would not be surprising if these numbers were somehow
related to the non-trivial topology of the QCD vacuum in any dimensions, though
by themselves they characterize its dynamical structure.

It is worth emphasizing that the structure of the severe IR
singularities in Euclidean space is much simpler than in Minkowski
space, where kinematical (un-physical) singularities due to the
light cone also exist \cite{14} (in this connection let us
remind that in Euclidean metric $q^2 = 0$ implies $q_i=0$ and
vice-versa, while in Minkowski metric this is not so). In this
case it is rather difficult to untangle them correctly from the
dynamical singularities, the only ones which are important for the
calculation of any physical observable from first principles. The consideration is
much more complicated in the configuration space \cite{14}. That is
why we prefer to work in the momentum space (where
propagators do not depend explicitly on the number of dimensions)
with Euclidean signature. We also prefer to work in the covariant
gauges in order to avoid peculiarities of the non-covariant gauges~\cite{17}, for example, how to untangle the gauge pole from
the dynamical one.

In principle, none of the regularization schemes (how to introduce
the IR regularization parameter in order to correctly treat the severe IR
singularities and thus to put them under control) should be
introduced by hand. Some of them have been cited in~\cite{9}.
However, not the regularization only is important but also its combination with the distribution theory.
Just this provides an adequate mathematical
framework for the correct treatment of all the Green's functions in QCD. 

The regularization of the INP IR singularities in four-dimensional QCD is determined
by the Laurent expansion (5.9) at $n=4$ as follows:

\begin{equation}
(q^2)^{- 2 - k } = { 1 \over \epsilon} a(k)[\delta^4(q)]^{(k)} +
f.t. = { 1 \over \epsilon} \Bigr[ a(k)[\delta^4(q)]^{(k)} +
O_k(\epsilon) \Bigl], \quad \epsilon \rightarrow 0^+,
\end{equation}
where $a(k)= \pi^2 / 2^{2k} k! \Gamma(2+k)$ is a finite constant depending only on $k$ and $[\delta^4(q)]^{(k)}$ represents
the $k$'s derivative of the $\delta$-function, see eqs.~(5.5)-(5.6). We point out once more that after introducing this expansion
everywhere one can fix the number of dimensions, i.e., put $d=n=4$ for QCD without any further problems. Indeed there will be no
other severe IR singularities with respect to $\epsilon$ as it goes to zero, but those explicitly shown in this expansion.

The finite terms in the DRE (5.10) is nothing else but its regular part, and it is

\begin{equation}
f.t. =(q^2)^{- 2 - k }_{-} + \epsilon (q^2)^{- 2 - k }_{-} \ln q^2 + O(\epsilon^2),
\end{equation}
where the functional $(q^2)^{- 2 - k }_{-}$ determines the number of the above-mentioned subtractions, see~\cite{3}.
These terms, however, do not undermine the INP MP IR renormalization program which will be
discussed below. The DRE (5.10) takes place only in four-dimensional QCD with
Euclidean signature. In other dimensions and/or Minkowski
signature it is much more complicated as pointed out above. As it
follows from this expansion any power-type INP IR singularity,
including the simplest one at $k=0$, scales as $1 /\epsilon$ as it
goes to zero. Just this plays a crucial role in the INP IR
renormalization of the theory within our approach. Such
kind of the DRE (5.10) does not exist for the PT IR singularity $(q^2)^{-1}$. 

In summary, first we have emphasized the
distribution nature of the INP (severe) IR singularities. Secondly, we have
explicitly shown how the DRM should be correctly and in a self-consistent way implemented into the distribution theory.
This makes it possible to put the severe IR singularities of any kind under a firm mathematical control.

\hspace{2mm}

\section{INP MP IR renormalization program}

\hspace{2mm}

Before starting to formulate the INP MP IR renormalization program to make theory finite when the IR dimensional regularization
parameter $\epsilon$ goes to zero
at final stage, let us remind the two important results obtained in our investigation. First, in the DRE (5.10) the pole over $\epsilon$ is always a simple pole  $1 / \epsilon$, and does depend on the number $k$ in its right-hand-side. It is easy to understand that just this makes the theory IR renormalizable; otherwise it would be non-renormalizable. The second point is that due to the Picard theorem only the simplest INP IR singularity at $k=0$ survives in the theory
when $q^2 \rightarrow 0$. Keeping in mind these two important observations, it is instructive to present the full gluon propagator
(2.14) and (2.16), on account of eq.~(3.1), as follows:

\begin{equation}
D_{\mu\nu}(q) =   {  i \over (q^2)^2} \Bigl[ T_{\mu\nu}(q) \Delta^2 L(q^2) + q^2 R_{\mu\nu}(q) \Bigr], \quad q^2 \in [0, \infty),
\end{equation}
where

\begin{equation}
R_{\mu\nu}(q) = T_{\mu\nu} d^{PT}(q^2) +  L_{\mu\nu}(q) { q^2 \tilde \lambda^{-1} \over q^2 + \tilde \lambda^{-1} \Delta^2},
\end{equation}
and we already introduced the renormalized version of the gauge-fixing parameter $\tilde \lambda^{-1} = Z_{\Delta} \lambda^{-1}$ and the mass gap itself $\Delta^2$. 
The superscript $'R'$, which has to stand for the renormalized gluon propagators in (6.1) is omitted, for simplicity.
$R_{\mu\nu}(q)$ is a regular function of $q^2$ when it goes to zero, while the purely longitudinal component of the full gluon propagator
(6.1) is strongly suppressed in this limit. In other words, the full gluon propagator (6.1) becomes purely
transverse in the $q^2 \rightarrow 0$ limit, i.e., at large distances.

The next step is to explicitly write down the DRE (5.10) for the simplest NP IR singularity at $k=0$

\begin{equation}
(q^2)^{- 2} = { 1 \over \epsilon} a(0)[\delta^4(q)] + f.t. = { 1 \over \epsilon} \Bigr[ a(0) \delta^4(q) +
O(\epsilon) \Bigl], \quad \epsilon \rightarrow 0^+,
\end{equation}
where $a(0) = \pi^2$. Due to the $\delta^4(q)$-function in the residue of this DRE, all the functions which appear under corresponding
skeleton loop integrals should finally be replaced by their values at $q=0$. Substituting this expansion into the full gluon
propagator (6.1), and after doing some algebra, one obtains

\begin{equation}
D_{\mu\nu}(q) = {1 \over \epsilon} \Bigl[ i T_{\mu\nu}(q) \pi^2 \Delta^2 Z_s \delta^4(q)+ O_{\mu\nu}(q; \epsilon) \Bigr], \quad \epsilon \rightarrow 0^+,
\end{equation}
which is nothing else but the DRE for the full gluon propagator (6.1), and we replace $L(q^2)$ by the constant $Z_s$ in accordance with relations (3.4)-(3.7). 
Here $O_{\mu\nu}(q; \epsilon)$ denotes the tensor terms of the $\epsilon$-order.

As it follows from the general DRE (5.10), whatever the severe IR singularity is in terms of $q^2$, in terms of the IR
regularization parameter $\epsilon$ it always scales as  $1 / \epsilon$. The IR renormalized full gluon propagator $D^R_{\mu\nu}(q)$,
which should exist or vanish as $\epsilon \rightarrow 0^+$, is to be defined as follows:

\begin{equation}
D^R_{\mu\nu}(q) = X(\epsilon) D_{\mu\nu}(q),
\end{equation}
and $X(\epsilon)$ is the INP MP IR renormalization constant for the full gluon propagator. Its general expansion in terms of the positive powers in $\epsilon$ is

\begin{equation}
 X(\epsilon) = \epsilon + O(\epsilon^2), \quad \epsilon \rightarrow 0^+.
\end{equation}
Evidently, such expansion for  IR renormalization constant makes the full gluon propagator (6.4) finite in the final $\epsilon \rightarrow 0^+$ limit. 

Thus, for the IR renormalized full gluon propagator (6.5), {\bf when $q^2$ is an independent loop variable}, one finally arrives at ($\epsilon \rightarrow 0^+$)

\begin{equation}
D^R_{\mu\nu}(q) = i T_{\mu\nu}(q) \pi^2 \Delta^2_{INP} \delta^4(q)+ \tilde{O}_{\mu\nu}(q; \epsilon),
\end{equation}
on account of eq.~(6.4), and where we introduce the notation (3.6). Here $\tilde{O}_{\mu\nu}(q; \epsilon)$
denotes the new tensor terms of the $\epsilon$-order, while the tensor terms of the $\epsilon^2$-order are omitted, for convenience.

However, this is not the whole story yet. In the multi-loop skeleton diagrams the more severe IR singularities than the simplest
INP IR singularity may appear. Then the product of at least the two $\delta$-functions at the same point becomes possible, if directly
substituting eq.~(6.7) for the different full gluon propagators. Such kind of the product is ill-defined in the theory of
distributions \cite{14}. In order to avoid this problem, the IR renormalized full gluon propagator (6.5) should be used in its general form as follows:

\begin{equation}
D^R_{\mu\nu}(q) = X(\epsilon) \times {  i \over (q^2)^2} \Bigl[ T_{\mu\nu}(q) \Delta^2 L(q^2) + q^2 R_{\mu\nu}(q) \Bigr],  \quad \epsilon \rightarrow 0^+.
\end{equation}
It is easy to see that if $q^2$ is independent loop variable then applying the DRE (5.10) for $k=0$, i.e., eq.~(6.3), one again arrives
at expression (6.7), on account of expansion (6.6).
Using this general expression for the IR renormalized full gluon propagator in the multi-loop skeleton diagrams instead of the $\delta$-functions
in the residues, their derivatives will appear according to the DRE (5.10). How to treat the loop integrals with the derivatives of the $\delta$-functions
is well-known procedure in the theory of the generalized functions (distributions)~\cite{14} (these integrals should be complemented by the corresponding
number of the subtractions at non-zero number $k$). The careful separation of the severely singular terms from the regular ones in the multi-loop skeleton diagrams is also important (in a single full gluon propagator this is automatically achieved, see eq.~(6.8)). However, as emphasized above, the IR renormalization of the theory will not be undermined
in this case as well, since a pole in $\epsilon$ is always a simple pole $1 / \epsilon$ whatever INP IR singularity with respect to $q^2$ appears in the theory, see again the DRE (5.10).

Expression (6.8) for the IR renormalized full gluon propagator does not depend whether the gluon
momentum is the independent skeleton loop variable or not. Indeed, if it approaches zero as a loop variable then expression (6.7) is to be
used, as described above. At the same time, if the gluon momentum {\bf is a free particle variable, i.e., it is an external ('leg') gluon momentum}, 
then expression (6.8) cannot be treated
as the distribution. So that it is a standard function and the DRE (6.3) is not the case to be used.
Then eq.~(6.8) behaves like

\begin{equation}
D^R_{\mu\nu}(q) = X(\epsilon) \times  {  i \over (q^2)^2} \Bigl[ T_{\mu\nu}(q) \Delta^2 L(q^2) + q^2 R_{\mu\nu}(q) \Bigr]  \sim \epsilon, \ \epsilon \rightarrow 0^+,
\end{equation}
again on account of expansion 6.6), i.e., it disappears in this limit.
This behaviour is gauge-invariant, does not depend on any truncations, approximations, or assumptions, based on a firm mathematical grounds, and thus it is a general one. 
It prevents the transverse full ("dressed") gluons from appear in asymptotic states (large distances) as physical
particles. So colour gluons can never be isolated. 

As pointed out in our book~\cite{9} (see discussion and references therein), we came back to the old IR slavery mechanism of gluon confinement but on a new basis 
(''new is well-forgotten old''). Let us note that the formulation of the $\epsilon \rightarrow 0$ intuitive mechanism for confinement has to be traced back to the 
pioneer papers by Pagels~\cite{18,19}. Here we put this mechanism on a firm mathematical ground in order to control severe IR singularities in QCD due to the
self-interaction of the massless gluon modes. In the past the IR slavery mechanism has been abandoned just because nobody knew how to deal with
such severe IR singularities in the correct mathematical way.  

There is also exists the possibility when $q^2$ is not a loop variable but it is internal one, so that the gluon confinement is not violated,
in any case. For example, such situation appears when one is going to calculate the potential between heavy quarks due to the exchange of the gluon.
We will provide such type of calculations in detail in appendix A.

\hspace{2mm}

\section{The mass gap and asymptotic freedom}

\hspace{2mm}

Let us now complete the investigation of the asymptotic properties of the full gluon propagator (1.1). We are left with its PT $q^2 \rightarrow \infty$ limit. 
The invariant function of the full gluon propagator or, equivalently, the running effective charge, is given in eq.~(1.2), namely

\begin{equation}
d(q^2; d) = { 1 \over 1 + \Pi(q^2;d) + (\Delta^2_t(d) / q^2)},
\end{equation}
where, in comparison with eq.~(1.2) we have already replaced $D$ by $d$, for further convenience.  
The invariant function $\Pi(q^2; d)$, being regular function at small $q^2$, may have the logarithmic divergences only in the PT $q^2 \rightarrow \infty$ limit.
Thus, one can neglect the ratio $(\Delta^2_t(d) / q^2)$ in eq.~(7.1), indeed.
Let us also put for further convenience $d(q^2; d) =  \alpha_s(q^2; \Lambda^2) / \alpha_s(\lambda)$, where $\Lambda^2$ and $\lambda$ are the UV regulating parameter and its dimensionless counterpart, respectively, while both $\alpha_s(q^2; \Lambda^2)$ and $\alpha_s(\lambda)$ present the corresponding running effective charges in the PT limit. Then for them from eq.~(7.1), one obtains

\begin{equation}
\alpha_s(q^2; \Lambda^2) = { \alpha_s(\lambda) \over 1 + b_0 \alpha_s(\lambda) \ln (q^2 / \Lambda^2)},
\end{equation}
where, as usual, we factorized the dependence on $b_0$ and $\alpha_s(\lambda)$ in the logarithmic divergent invariant function calculated to leading order (at one loop level).
This is nothing else but the sum of the main PT logarithms~\cite{1,2,20,21,22}. Here $b_0= 11/4 \pi > 0$ is the colour group factor for the Yang-Mills (YM) fields.
Note, we do not need to calculate the Gell-Mann-Low  function as a renormalizable group equations solution in powers of the coupling constant ~\cite{20} in order to get finally to the expression (7.2).
It has been already done within our approach. 
Evidently, in this regularized expression one can not go directly to the $(\Lambda^2, \lambda \rightarrow \infty)$ limits in order to finally arrive at the renormalized expression. 
Some intermediate steps are to be performed in order to complete the corresponding renormalization program.

Due to the Picard effect discussed above, we already know that there exists the finite mass squared parameter defined in eq.~(3.7), namely $\Delta^2_{AF} = \Delta^2 \times Z_r$,
which survives the PT limit, see eqs.~(3.5) and (3.7). It is basically determined by the renormalized version of the tadpole term, i.e., a mass gap $\Delta^2$ itself, which dominates the dynamical structure
of the full gluon propagator in the IR region.
This effect explicitly shows that in some cases it is useful to distinguish between the suppression of the
contribution $(\Delta^2_t(d) / q^2)$ in the PT limit, i.e., the term itself can be suppressed/neglected but the constant itself survives and may become finite.
It is possible to say that the Picard number $Z_r$ (3.7) ''pulls out'' the mass gap $\Delta^2$ from the NP region of small gluon momenta to the PT region of large gluon momenta. 
All this means that within our approach the UV regulating parameter $\Lambda^2$ can be always replaced by

\begin{equation}
\Lambda^2 = f(\lambda) \Delta^2_{AF},
\end{equation}
where $f(\lambda)$ is the corresponding dimensionless function, which is quadratically divergent in the $\lambda \rightarrow \infty$ limit, but otherwise
remains arbitrary. Let us emphasize that such kind of the relation makes sense to use only because we know in advance that the finite constant $\Delta^2_{AF}$ exists and comes out into the play in the PT limit, see eq.~(3.7). It is neither arbitrary nor introduced by hand, i.e., it is not only formally mathematically correct, but also being physically meaningful within the mass gap approach to QCD. 
Substituting this relation into eq.~(7.2) and doing some algebra, one obtains 

\begin{equation}
\alpha_s(q^2) = { \alpha_s \over 1 + b_0 \alpha_s \ln (q^2 / \Delta^2_{AF})},
\end{equation}
if and only if

\begin{equation}
\alpha_s = { \alpha_s(\lambda) \over 1 - b_0 \alpha_s(\lambda) \ln f(\lambda)}
\end{equation}
exists and is finite in the weak coupling limit $\alpha_s(\lambda) \rightarrow 0$ when $\lambda \rightarrow \infty$.
The finite $\alpha_s$ can be identified with the fine structure constant of the strong interactions, calculated at some fixed scale,
$\alpha_s \equiv \alpha_s(M_Z) = 0.1184$~\cite{21}.
The existence and finiteness of $\alpha_s$ can be shown within the standard renormalization group equations arguments. 
Indeed, from eq.~(7.5) it follows

\begin{equation}
\ln f(\lambda) = { \alpha_s - \alpha_s(\lambda) \over  \alpha_s b_0 \alpha_s(\lambda)} \rightarrow { 1 \over b_0 \alpha_s(\lambda)},
\quad \alpha (\lambda) \rightarrow 0, \quad \lambda \rightarrow \infty,
\end{equation}
which means that

\begin{equation}
f(\lambda) = \exp ( 1 / b_0 \alpha_s(\lambda))
\end{equation}
in the weak coupling limit shown in eq.~(7.6). Substituting this into relation (7.3), it becomes

\begin{equation}
\lim_{(\Lambda^2, \lambda) \rightarrow \infty} \Lambda^2 \exp \left[ - { 1 \over b_0 \alpha_s(\lambda)} \right] = \Delta^2_{AF},
\quad \alpha(\lambda) \rightarrow 0,
\end{equation}
which is the finite limit of the renormalization group equations solution~\cite{22}. Note that with $\lambda \rightarrow \infty$ the effective charge
$\alpha_s(\lambda)$ has to have the logarithmic fall off (weak coupling limit) in order to satisfy to the final limit in eq.~(7.8).
It is easy to check that such behaviour justifies the existence of the finite $\alpha_s$ in eq.~(7.5) as well.

At very large gluon momentum $q^2$ from eq.~(7.4) one recovers

\begin{equation}
\alpha_s(q^2) = { 1 \over b_0 \ln (q^2 / \Lambda^2_{YM})} \rightarrow 0, \quad q^2 \rightarrow \infty,
\end{equation}
which is AF famous formula~\cite{1,2,4, 20, 22}. In order to get this expression from eq.~(7.4) we replace $\Delta^2_{AF}$ there by $\Lambda^2_{YM}$  for the purely YM fields. 
Evidently, this can be always done by redefining the dimensionless function $f(\lambda)$ in eq.~(7.3), since
$\Lambda^2 = f(\lambda) \Delta^2_{AF}= f(\lambda) A^{-1}A \Lambda^2_{YM}= \tilde{f}(\lambda) \Lambda^2_{YM}$, and $A$ is the arbitrary finite number. It disappears from the final expressions.

Going now back to eq.~(1.1), on account of eq.~(7.4), one obtains 

\begin{equation}
D_{\mu\nu}(q) \sim iT_{\mu\nu}(q) \Bigl[ { \alpha_s \over  1 + \alpha_s b_0 \ln(q^2/ \Lambda^2_{YM}) } \Bigr] (1 / q^2) 
\end{equation}
in the PT $q^2 \rightarrow \infty$ limit and the longitudinal component is not shown, for simplicity. It is necessary to stress that this expression has been obtained 
on the basis of the asymptotic properties of the singular solution investigated in the previous sections. So that the singular solution possesses  the AF behaviour.
The explicit presence of the mass gap in the full gluon propagator
makes it possible to explain the logarithmic scale breaking in the PT regime. The mass gap coming from the IR region, nevertheless, survives the PT
renormalization program and finally becomes $\Lambda^2_{YM}$ in the weak coupling limit as described above.
$\Lambda^2_{YM}$ determines the scale of the non-trivial PT dynamical structure of the YM ground state. 

Let us remind that the mass scale parameter $\Lambda^2_{YM}$ cannot be generated
by the PT itself. Due to the renormalization group equations arguments \cite{20,23}, any mass to which can be assigned some physical meaning disappears
according to
\begin{equation}
M  \sim \mu \exp(- 1 /  b_0 g^2), \quad \quad g^2 \rightarrow 0,
\end{equation}
where $\mu$ is the arbitrary renormalization point and $b_0 > 0$. In other words, in every order of the PT it is exponentially suppressed.
So that the question where the scale $\Lambda^2_{YM}$ in eq.~(7.10) comes from? is not answered by the PT itself.
None a finite mass can survive in the PT weak coupling limit or, equivalently, in the PT $q^2 \rightarrow \infty$ regime, indeed.
The mass gap approach solves this problem, namely the mystery of the scale breaking (which is completely NP effect) in the AF regime.
Such kind of the scale parameter has to come from the IR region being of the NP dynamical origin, as described above.

All this is a manifestation that "the problems encountered in perturbation theory are not mere mathematical artefacts but rather signify
deep properties of the full theory"~\cite{24}. The message that we are trying to convey is that the INP dynamical structure of the full gluon propagator
indicates the existence of its non-trivial perturbative  structure and {\it vice \ versa}. So that our solution coincides with the PT QCD solution in the PT limit~\cite{1,2}, as it has to be, since the QCD
by itself is  a renormalizable quantum field theory.The difference is that our solution explains where  $\Lambda^2_{YM}$ comes from, while the PT QCD cannot.

\hspace{2mm}

\section{Discussion}

\hspace{2mm}

Let us now discuss the connection of our advance results to the Elitzur's theorem~\cite{25,26} (in the second paper one can find also review of the present status of the confinement problem).
Theorem itself is read as follows: 
   
\hspace{0.1mm}

{\bf Elitzur's theorem:} A spontaneous breaking of local symmetry for a symmetrical gauge theory without gauge fixing is impossible. 

\hspace{0.1mm}   
   
It has been proven in a lattice field theory for the abelian system like Quantum Electrodynamics (QED), and is supposed to be valid in continuum field theory. From our consideration it follows that
this theorem is valid for the continuous QED (omitting the YM part here, one obtains, in fact, such a theory). Obviously, it is valid 
for the PT QCD theory as well. This is so indeed, if to understand this theorem as the request that the gauge symmetry of the QCD Lagrangian and its ground state should coincide.
However, to the mass gap approach to QCD this theorem cannot be applied since 
the above-mentioned gauge symmetries do not coincide from the very beginning.
Evidently, at the moment of the formulation of his theorem, Elitzur has had no idea about the complicated gauge and dynamical structures of the QCD true ground state. 
Nevertheless, this theorem should be taken into account, since the PT QCD full gluon propagator is independent solution of the initial gluon SD equation~\cite{3}.  
That is a reason why one can formally register the four different ''flavours/species'' of the gluon
field configurations in the true QCD vacuum: the free gluons, the PT QCD gluons for which the local gauge symmetry of the Lagrangian and its ground state  is always maintained. 
But as it follows from many studies (including ours) the free and the PT QCD gluon propagators do not satisfy to the confinement boundary condition at low energies (large distances).
The two other our solutions-- for massive gluons~\cite{15}  and singular gluons,
dynamically originated from the vacuum, which gauge structure does not coincide with the Lagrangian gauge symmetry. However, as underlined in our first paper this non-coincidence is not so important. 
The important thing is that the gauge-invariant solution at least for the one of the corresponding equations of motion 
has to exist in order to calculate the physical observables from first principles in QCD.  Just the singular solution is a such one, as emphasized above.  
It satisfies both boundary conditions, while gluons remaining massless, i.e., the gauge symmetry is not violated for it. Moreover, by gauge fixing we understand the choice of the numerical value 
for $\lambda^{-1}$ in the initial ST identity (1.3) and not the generalized gauge $\xi(q^2)$ itself. So that if the gauge is not fixed then the symmetry of the ground state can be broken as it follows
from our understanding of this theorem. However, whether we correctly understand or not Elitzur's theorem, contrary to the JW's theorem in any case, it has a little physical influence on the mass 
gap approach to QCD.

Let us also point out that the singular solution is in a fair agreement with the {\bf Strocchi's theorem}~\cite{27}, consisting of the five Propositions and their Proofs 
(so that it is difficult to explicitly present them here). It admits such a singular behaviour for the full gluon propagator in the YM-type theory. Then the quark-antiquark potential 
may behave at infinity like $r^N$, $N$ positive. In fact, we proved that $N=1$ due to the Picard's effect (see appendix A).

\hspace{1mm}

\section{Conclusions}

\hspace{1mm}

Since this paper is the last one of our series, let us make a few general things perfectly clear. In the introduction to the first paper~\cite{3} the two  important questions have been raised: 
how does a massless theory generate a mass? and why coloured objects--gluons and quarks--have been not observed up to present days? The existing QCD 
theory, i.e., the PT QCD is unable to answer these existential questions.
The essence of our mass gap approach to QCD is the exactly derived constraint on any solution to QCD~\cite{3}. It makes it possible to derive a new expression for the full gluon propagator (1.1),
directly depending on the tadpole term,  and being nothing else but the NL transcendental equation. Its solution allowed  to answer these important questions. Let us begin with a mass generation problem.

\hspace{2mm}

\subsection{Origin of a mass}

\hspace{2mm}

The self-interaction of the multiplying massless gluon modes leads to the generation of the 
infinite number of the quantities having the dimensions of a mass squared in the QCD ground state (see Fig. 1 and its description in the text of~\cite{3}) . However, only one of 
these combinations, namely the tadpole term generates a mass squared to which can be assigned a physical meaning after the corresponding renormalization program has been performed. 
The gauge symmetry of the QCD Lagrangian is spontaneously broken in its ground state by the explicit presence of this term.
It is dynamically generated by the self-interaction 
of the multiplying massless gluon modes, involving the point-like four-gluon vertex only~\cite{6,28}, i.e., it is a fundamental quantity (''mass without mass''~\cite{29}), so that it cannot be a bound 
state of anything. This contribution to the full gluon self-energy does not depend on the external gluon momentum, and thus generates the scale parameter, having the dimension of a mass squared~\cite{30}, indeed. Its renormalization version has been conventionally called a mass gap in accordance with the question asked above and sheds light on understanding ''one of the deep unsolved physics 
mysteries, the mass gap''~\cite{6}. The QCD is a self-consistent quantum gauge field theory. It does not need any 
extra degrees of freedom (such as Higgs fields~\cite{31}) or to extract a mass by some other ways in the QCD vacuum.
The mass gap cannot be disregarded from the theory and its ground state by any means, but the PT renormalizability of QCD will not be affected. 
The tadpole-like terms $\Delta^2_q$ and $\Delta^2_g$ cannot be considered as sources of the mass gap. They appear as a result of the subtraction scheme formulated  
in~\cite{3} (the initial skeleton loop integrals depend on the external gluon momentum). These terms finally have to be removed from the theory.

In order to get the singular solution, which satisfies both boundary conditions, we replaced initial NL transcendental equation (2.2) by the corresponding expansion not in powers of the coupling constant
but in powers of the mass gap. Such expansion has been called the cluster one in~\cite{6}. So that in the presence of the mass gap 
the role of the QCD coupling constant $g^2$ becomes unimportant. This is also evidence of the ''dimensional transmutation'',
$g^2 \rightarrow \Delta^2$~\cite{20,23}, which occurs whenever a massless theory acquires mass dynamically. It is a general feature of spontaneously
symmetry breaking in field theories. Both new solutions massive and singular reflect this evidence, while the previous solution of the PT QCD does not.

Concluding finally this part, let us once more underline that only one special type of the self-interaction of the multiplying massless gluon modes (described above) dynamically generates a mass in the QCD. 
Precisely it allows one to extend of the JW mass gap  conception~\cite{6} to the true QCD ground state, as it has been done within the mass gap approach to QCD in our paper series.

\hspace{2mm}

\subsection{Gluon confinement}

\hspace{2mm}

The gauge symmetry of the QCD Lagrangian is not a symmetry of its ground state, which has been proven in our first paper~\cite{3}. In this paper we have 
found the confining singular solution for the full gluon propagator, which, nevertheless, remaining massless. In the presence of the mass gap 
all the possible severe IR singularities (more singular than $~1/q^2$ of the free and PT QCD  gluon states), which generated by the self-interaction of the multiplying massless gluon modes,  
can be accommodated into the single full gluon propagator. 
Possessing such a singular behaviour, it suppresses all the above-mentioned gluon states to appear in the physical spectrum at large distances ($ q^2 \rightarrow 0$).
It is necessary to understand that without mass gap there exists no any other mechanism for a such suppression. To correctly treat the severe IR singularities,
we developed a formalism of the joint use of the theory of functions of complex variables, and the DRM implemented  into the theory 
of distributions in a self-consistent way (for the first time to our best knowledge). As a consequence, the dependence on the IR dimensional regularization parameter $\epsilon \rightarrow 0$ 
final limit appears in the theory. If the gluon momentum is independent loop variable then the full gluon propagator (6.7) has a finite limit. It always possesses only the ''physical'' transverse 
degrees of freedom and gluons always remain massless as well. This means that gauge symmetry is not violated by this solution even in the explicit presence of the mass gap. Moreover, due its presence our solution possesses the scale breaking of logarithmic-type in the AF regime at large gluon momentum and power-type scale breaking at very small gluon momentum.
Finally this effect leads to its confinement to the true QCD vacuum within our $\epsilon \rightarrow 0$-formalism. A big advantage of this formalism is that in the case of the finite result 
in eq.~(6.7) it  automatically becomes free from all the types of the PT ''contaminations'' (they become terms of $\tilde{O}_{\mu\nu}(q; \epsilon)$-order, and thus vanish in the final 
$\epsilon \rightarrow 0$ limit, see again eq.~(6.7))

At the finite gluon momentum the full gluon propagator (1.1) generates exactly defined the gluon pole mass~\cite{15}. However, it cannot be treated as the mass-shell singularity. This means that
acquiring mass dynamically, the massive gluon, nevertheless, cannot appear in the physical spectrum.  Though confining, this solution at large gluon momentum goes to the free gluon propagator, i.e., 
it does not possess the AF regime, and thus does not satisfy one of the necessary boundary conditions mentioned above.

\hspace{2mm}

\section{Summary}

\hspace{2mm}

The mass gap approach to QCD and its true ground state shows that confinement of gluons is absolute~\cite{20}, indeed, i.e., colour gluons may exist in 
the vacuum or inside hadrons only.The gluon confinement (which is the first to be solved) along with the quark confinement problems both are parts of a long-standing general 
colour confinement problem in QCD. The two existential  questions of a mass dynamical generation 
and colour gluons confinement, as it turned out to be in close connection with each other, have been answered in our investigations through out the three paper series. 
The two new confining solutions for the full gluon propagator have been discovered. 
The massive solution is confining, i.e., no the mass-shell singularity at some finite point, but it is not satisfying the boundary condition of the AF regime at short distances. 
The singular solution is also confining and possessing AF regime as well, i.e., satisfying both boundary conditions. It always depends on the transverse (''physical'') degrees of freedom
and gluons also always remain massless. The mass gap approach to QCD provides the scale $\Delta^2_{INP}$ which can be responsible for the first confinement phase transition in QCD. 
It is based on the mass gap $\Delta^2$ itself, see eq.~(3.6). Precisely this solution will make it possible to calculate the physical observables in QCD from first principles, which is ultimate goal of this theory.
 
The mass gap approach to QCD or, equivalently, the confining QCD theory opens a bright perspectives for further work in the gluon dynamics sector in QCD (see appendix B).  
However, going beyond the YM sector is the most important but more difficult direction for future work.
To solve the quark SD equation of motion by the substitution into it the confining solution for the full gluon propagator will make it possible to solve the general colour confinement problem
in QCD. This will happen only in the case of establishing confining solution for the quark Green's full function. It would mean that the full quark propagator could not be put on the mass-shell, i.e., 
it has to be an entire function of its argument~\cite{20}. However, preliminary solution of the gluon confinement  problem (already done in our series) and only after will come solution 
of the quark confinement problem and {\it not \ vice \ versa}, as emphasized just above.

\section*{acknowledgments}

The authors are grateful to P. Forg\'{a}cs, J. Nyiri, T.S. Bir\'{o}, M. Vas\'{u}th, Gy. Kluge
for useful suggestions, remarks, discussions and help. The work was supported by the Hungarian National Research, Development and Innovation Office (NKFIH) under 
the contract numbers OTKA K135515, K123815 and NKFIH 2019-2.1.11-T\'ET-2019-00078, 2019-2.1.11-T\'ET-2019-00050, the Wigner GPU Laboratory and THOR Cost Action CA15213.

\appendix

\hspace{2mm}

\section{The linear rising potential}

\hspace{2mm}

From the initial general expression (6.1) the corresponding effective charge associated with its transverse component is to be defined as follows:

\begin{equation}
\alpha(q^2; \Delta^2) = { \Delta^2 \over q^2} \times L(q^2) + d^{PT}(q^2),
\end{equation}
where the Laurent expansion is defined in eq.~(3.2), and taking into account eq.~(6.2). However, due to the Picard theorem, discussed in detail in section 3, at large distances
($q^2 \rightarrow 0$) limit in the full gluon propagator (6.1) survives only the simplest severe IR singularity $\sim 1/q^4$, on account of the definition (3.6). The  $d^{PT}(q^2)$ is regular function of its argument and it is left here in order to reproduce the Cornell potential below. So that the renormalized effective charge to leading order at large distance becomes

\begin{equation}
\alpha(q^2; \Delta_{INP}^2) = { \Delta_{INP}^2 \over q^2},
\end{equation}
where $\Delta^2_{INP} = \Delta^2 Z_s$ in complete agreement with the notations introduced in section 3.
For the phenomenological applications considered in this appendix and in the next one, eq.~(A.2) is fully justified to use.

The potential between heavy quarks is defined as follows:

\begin{equation}
V(r) = - 4 \pi C_2(R) \int { d^3 q \over (2 \pi)^3} e^{i \bar{q} \bar{r}} { \alpha(q^2; \Delta^2_{INP}) \over q^2},
\end{equation}
where $C_2(R) = (N_c^2 -1)/ 2N_c = 4/3$ is the positive eigenvalue of the quadratic Casimir operators for the fundamental representation of $SU(3)$.

The general expression for the power-type effective charges is~\cite{32}

\begin{equation}
\int { d^n q \over (2 \pi)^n} e^{iqr} { 1 \over (q^2)^{\lambda}} = { \Gamma(n/2 - \lambda) \over 4^{\lambda} \pi^{n/2}
\Gamma(\lambda)} (r^2)^{\lambda - n/2}.
\end{equation}
Substituting further eq.~(A.2) into eq.~(A.3) and using this expression for $\lambda =2, \ n=3$ and $\Gamma(-1/2)= - 2 \sqrt{\pi}$,
one finally obtains

\begin{equation}
V^{INP}(r) = { 2 \over 3} \Delta^2_{INP} \ r = \sigma \ r, \quad r \rightarrow \infty,
\end{equation}
and $\sigma$ is the string tension. The relation between the scale of the INP QCD dynamics $\Delta^2_{INP}$ and the string tension $\sigma$ then becomes

\begin{equation}
\Delta^2_{INP} = { 3 \over 2} \sigma.
\end{equation}

From lattice simulations~\cite{21,33,34} it is well-known that

\begin{equation}
\sqrt{\sigma} = 0.420 \ GeV, \quad \sigma = 0.1764 \ GeV^2,
\end{equation}
so that for the scale $\Delta^2_{INP}$, one obtains the value as follows:

\begin{equation}
\Delta_{INP} = 0.5144 \ GeV, \quad  \Delta^2_{INP} = 0.2646 \ GeV^2.
\end{equation}
Of course, this value should not be accepted literally, since it depends on the string tension's value, but the relation (A.6) is exactly defined.

Is is instructive to calculate the potential for the PT QCD gluon propagator (2.17), which behaves as $\sim 1/q^2$ in the whole gluon momentum range.
This means that its effective charge at zero becomes simply constant, namely $d^{PT}(0) = \alpha^{PT}(0)$, which for the free gluon effective charge is $d^0(0) = \alpha^0(0)=1$. 
Then from the expressions (A.3) and (A.4) for $\lambda =1, \ n=3$ and $\Gamma(1/2)= \sqrt{\pi}$, one finally obtains the Coulomb potential, as expected

\begin{equation}
V^{PT}(r) = - { 4 \over 3} { \alpha_s \over r}, \quad r \rightarrow 0,
\end{equation}
where $\alpha_s  = \alpha^{PT}(0)$ and $\alpha_s$ itself is usually calculated at a scale of Z boson mass, i.e., $\alpha_s \equiv \alpha_s(M_Z) = 0.1184$~\cite{21}.

The Cornell potential, in accordance with eq.~(A.1), becomes

\begin{equation}
V(r) = V^{INP}(r) +V^{PT}(r) = { 2 \over 3} \Delta^2_{INP} \ r - { 4 \over 3} { \alpha_s \over r}.
\end{equation}
Evidently, this sum reflects our starting eq.~ (A.1). 
Let us point out that at finite gluon momenta, i.e., at small but finite distances, the potential is of Yukawa-type, namely
$V^{NP}(r) \sim (1/r) \exp(-M_gr)$, where $M_g$ is the exactly defined gluon pole mass~\cite{15}. So that at very large and very small distances
the Cornell potential (A.10) is dominated by its INP and PT contributions, respectively, while at the finite distances its NP contribution has to be taken
into account as well.

The Picard effect is a main ''culprit'' behind the phenomenon that the potential between heavy quarks is only linear confining one,
calculating by lattice QCD, as pointed out above, i.e., there is no any power-type confining potential at large distances.
Due to this effect only simplest INP IR singularity $\sim 1/ q^4$ survives at large distances ($q^2 \rightarrow 0$), while all others are suppressed (see subsection 3.1). This is the one of main phenomenological consequences which directly follows from the mass gap approach to the QCD ground state. If its gauge symmetry would have been the same as of the QCD Lagrangian, then there were no the linear rising potential at all, and only Coulomb-type one would have been existed.
In other words, the Picard effect, which shows up itself within the mass gap approach, explains why there is only linear rising potential between
heavy quarks in QCD.

And finally, to analyse the interaction between heavy quarks in terms of the confining linear rising potential is justified. In this
case the response of the vacuum (radiative corrections or, equivalently, vacuum fluctuations) can be neglected, so that the full quark-gluon vertex is well-approximated by its point-like counterpart. For light quarks this simplification does not work, since the response of the vacuum is strong, and the full quark-gluon vertices are to be taken into account. So the only way to understand confinement of light quarks is the investigation of analytical and asymptotic properties of the quarks Green's functions, i.e.,  their full propagators and the corresponding full vertices. They will be substantially
modified due to the strong vacuum response. It has been shown~\cite{35,36} that the ladder approximation (LA) is not self-consistent in QCD due to the strong interaction between colour charges. 
It can be used only for heavy quarks, as explained above, or in the AF regime, when the coupling constant becomes weak.

\section{The renormalized running effective charge and the $\beta$-function}

\begin{figure} [h!]
\begin{center}
\includegraphics[width=11.0truecm]{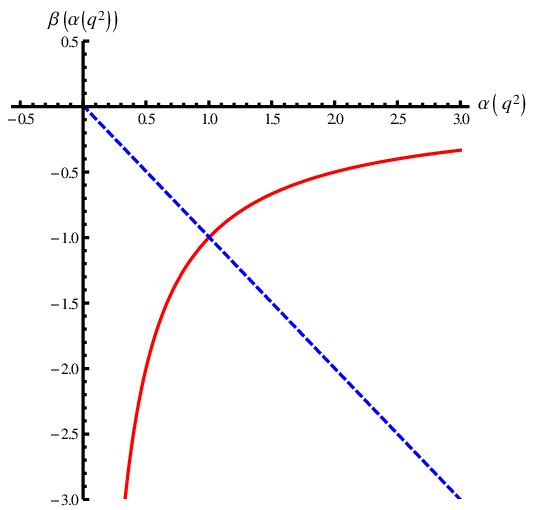}
\caption{ The straight dash line shows the $\beta$-function as a function of the effective charge $\alpha(q^2)$. The solid curve stands for it
as a function of $z$, see eq.~(B.3).}
\label{fig:2}
\end{center}
\end{figure}

\hspace{2mm}

The another interesting phenomenological consequence which directly follows from the mass gap approach to QCD at large
distances ($q^2 \rightarrow 0$) is the relation between the renormalized running effective charge (A.2)

\begin{equation}
\alpha(q^2; \Delta_{INP}^2) = { \Delta_{INP}^2 \over q^2}=  {1 \over z}, \quad z = {q^2 \over \Delta^2_{INP}}
\end{equation}
and the corresponding $\beta$-function. The renormalization group equation for the renormalized effective charge, which determines the corresponding
$\beta$-function, is as follows:

\begin{equation}
q^2 {d \alpha(q^2; \Delta^2_{INP}) \over dq^2} = \beta( \alpha(q^2; \Delta^2_{INP})).
\end{equation}
Then from eq.~(B.1) one immediately obtains

\begin{equation}
\beta( \alpha(q^2; \Delta^2_{INP})) = - \alpha(q^2; \Delta^2_{INP}) = - {1 \over z}.
\end{equation}

Thus, one can conclude that the corresponding $\beta$-function as a function of its arguments is always in the domain of attraction, i.e., it is always negative.
So it has no IR stable fixed point. Moreover, it scales independently from
the value of the mass gap $\Delta^2_{INP}$, see Fig. B.2 and eq.~(B.3). In other words, it possesses both features required for the confining theory~\cite{20}, indeed.
In phenomenological applications the $\beta$-function itself and the ratio

\begin{equation}
{\beta( \alpha(q^2; \Delta^2_{INP})) \over \alpha(q^2; \Delta^2_{INP})} = - 1,
\end{equation}
one needs to know in order to calculate the TNP quantities, which are free from all the types of the PT contributions ("contaminations", see eq.~(4.4)).
These are the above-mentioned YM part of the bag constant, the gluon condensate, the topological susceptibility etc.~\cite{9,37,38,39,40,41,42,43}
(and references therein).

In the previous appendix it was demonstrated that the growing effective charge (B.1) leads to the linear rising potential between heavy quarks
in complete agreement with lattice calculations. Indeed, there already exists direct lattice evidence of such kind behaviour of the running effective charge.
A NP finite-size scaling technics was used in~\cite{44} to study the evolution of the running coupling
in $SU(3)$ YM lattice theory. By using the two-loop $\beta$-function it is shown to evolve according to the PT at high energies, while at low energies it
is shown to grow. 

Concluding, it is worth noting, that in this appendix we have used from the very beginning all the expressions for the running effective charge and the corresponding $\beta$-function, which are free from 
all the types of the  PT "contaminations", as it needs be, indeed. In other words, necessary subtraction has been already done in eq.~(A.1) in accordance with eq.~(4.4), and we omitted the
superscript ''TNP'', for simplicity.

\section{The strong and short-ranged character of the gluon delta-function ''potential''}

In the previous appendixes the gluon momentum was not a loop variable in order to determine the linear rising potential between quarks and the corresponding
$\beta$-function. However, if the gluon momentum squared is a loop variable, then the gluon ''potential'' $D(q;\Delta^2_{INP})$ is to be defined as follows:

\begin{equation}
D(q; \Delta^2_{INP}) = \Delta^2_{INP} \pi^2  \delta^4(q),
\end{equation}
as it comes out from the expression (6.7). For the components of the $\delta^4(q)$-function it is convenient to use the following expression

\begin{equation}
\delta(q_i) = \lim_{ \Delta_{INP} \rightarrow 0} { 1 \over \Delta_{INP} \sqrt{\pi}} e^{-{q_i^2 \over \Delta^2_{INP}}}, \quad i=0,1,2,3.
\end{equation}
Then $D(q; \Delta^2_{INP})$ becomes

\begin{equation}
D(q; \Delta^2_{INP}) =\lim_{ \Delta^2_{INP} \rightarrow 0} { 1 \over \Delta^2_{INP} } e^{-{q^2 \over \Delta^2_{INP}}},
\end{equation}
and in the dimensionless units it is shown in Fig. C.3.

\begin{figure}[!h]
\begin{center}
\includegraphics[width=11.0truecm]{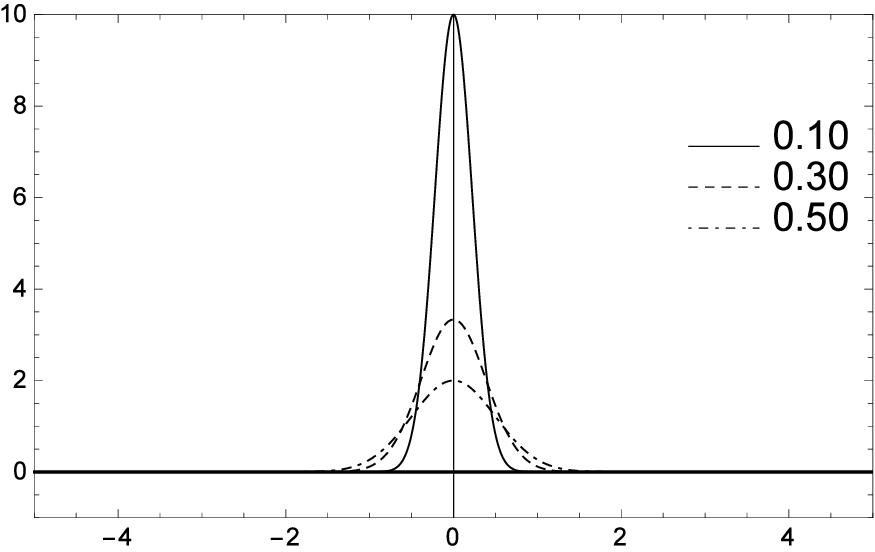}
\caption{The delta-function potential with the mass gap squared as the sequences of zero-centered normal distributions. In the upper right corner the values
of the mass gap squared used are shown but without its dimension.}
\end{center}
\end{figure}

The $\delta$-type gluon ''potential'' is strong and short-ranged, similar to the nuclear force. For the above-shown values for the INP mass gap the range
of the interaction in the God units $\hbar = c =1$ is about $ a fm \sim a \times 10^{-13} cm$ with constant $a < 1 $. The range of the nuclear forces
is also about $ a fm \sim a \times 10^{-13} cm$ but constant $a > 1 $. It is interesting to compare our estimate with the lattice range of the 
three-quark potential~\cite{45}, which is about $ 0.7 fm \sim 0.7 \times 10^{-13} cm$, and it is in a good agreement with our estimate.

JW in~\cite{6} explain that one needs a mass gap in order to demonstrate why the nuclear force is strong but short-ranged.
In this paper we have shown that this phenomenon takes place even at the fundamental gluon level, i.e., the interaction between objects by exchange of the
confining gluons, depending on the mass gap, is strong and short-ranged, indeed. They also pointed out that a correct theory of QCD must have "quark confinement".
Here we have shown that the mass gap approach needs to have gluon confinement. Let us underline once more (see conclusions in~\cite{3})  that the quark confinement problem can be solved
only on the basis of the solution of the gluon confinement one, present in this paper. It is beyond the scope of this work.

\section{The NL transcendental character of the tadpole term}

The constant tadpole term $\Pi_{\rho\sigma}^t(D)$ has been defined in~\cite{3} and in Euclidean metrics it is as follows:

\begin{equation}
\Pi_{\rho\sigma}^t(D) \sim  \int i d^4 l D_{\alpha\beta}(l) T^0_{\rho\sigma\alpha\beta} = \delta_{\rho\sigma} \Delta^2_t(D), 
\end{equation}
where and in what follows below all the overall numerical numbers will be omitted for simplicity, since they are not important for our primary purpose in this  appendix.
The renormalized version of the tadpole term is

\begin{equation}
\Delta^2_t(D)   = Z(\lambda) \times \Delta^2,
\end{equation}
where the NP renormalization constant has to behave like $Z(\lambda) \sim \lambda^2$ as $\lambda \rightarrow \infty$: otherwise remaining arbitrary (see below). 
Reminding that the mass gap $\Delta^2$ itself is finite and positive, by definition.

Substituting further the expression for the full gluon propagator (1.1) into eq.~(D.1) and doing some tensor algebra derivations, one finally obtains 

\begin{equation}
 Z(\lambda) \sim \int d^4 x  \left[ { 3 \over x^2 + x^2 \Pi(x^2) +  Z(\lambda) }  +{ \xi_0 \over x^2 + \xi_0  Z(\lambda)} \right],
\end{equation}
on account of the relation (D.2) and introducing the dimensionless variable $x=l / \Delta$. Here we came back to the definition $\xi_0 =\lambda^{-1}$ in order not to mixed up 
the gauge fixing parameter  $\lambda^{-1}$ with the dimensionless UV regulating parameter $\lambda$, denoted above.

Expression (D.3) for the NP renormalization constant definitely is the transcendental NL equation, and thus the tadpole itself. It cannot be identified with the UV cut-off $\Lambda^2 =\lambda^2 \Delta^2$.
In this connection, let us remind that such kind of the skeleton loop integrals are assumed to be regularized from above and below~\cite{3}.
Also it is easy to check that it really behaves as $Z(\lambda) \sim \lambda^2$ in the  $\lambda \rightarrow \infty$ limit (up to a possible logarithmic corrections coming from  the invariant function
$\Pi(x^2)$ (if any)). For this one needs to use that $d^4 x = x^3 dx = (1/2) x^2 d x^2$ and integration over variable $x^2$ is going from zero to $\lambda^2$ (numbers coming from the integration over
angular variables have been omitted as pointed out above).


{}

\end{document}